\documentclass[12pt,a4paper]{article}
\usepackage{epsf}
\usepackage{amsmath,amssymb}
\usepackage{url,cite,ifpdf,slashed,multirow,tabularx}

\ifpdf                                     
  \usepackage{graphicx}                    
\else                                      
  \usepackage[dvips]{graphicx}             
\fi

\begin{document}

\renewcommand{\thefootnote}{\fnsymbol{footnote}} 
\begin{titlepage}

\begin{center}

\hfill UT--13--37\\
\hfill IPMU--13--0203\\
\hfill October, 2013\\

\vskip .75in

{\Large \bf 
  Reconstructing Supersymmetric Contribution to \\[0.3em]
Muon Anomalous
  Magnetic Dipole
   Moment at ILC
}

\vskip .75in

{\large
  \textbf{Motoi Endo}$^{\rm (a,b)}$,
  \textbf{Koichi Hamaguchi}$^{\rm (a,b)}$,
  \textbf{Sho Iwamoto}$^{\rm (b)}$\footnote{Research Fellow of the Japan Society for the Promotion of Science},
  \\[0.3em]
  \textbf{Teppei Kitahara}$^{\rm (a)}$,
  and \textbf{Takeo Moroi}$^{\rm (a,b)}$}

\vskip 0.25in

$^{\rm (a)}${\em Department of Physics, University of Tokyo, Tokyo 113--0033, Japan}
\vskip 0.1in
$^{\rm (b)}${\em Kavli IPMU (WPI), University of Tokyo, Kashiwa, Chiba 277--8583, Japan}

\end{center}

\vskip .5in

\begin{abstract}

  We study the possibility to determine the supersymmetric (SUSY)
  contribution to the muon anomalous magnetic dipole moment by using
  ILC measurements of the properties of superparticles.  Assuming that
  the contribution is as large as the current discrepancy between the
  result of the Brookhaven E821 experiment and the standard-model
  prediction, we discuss how and how accurately the SUSY contribution
  can be reconstructed.  We will show that, in a sample point, the
  reconstruction can be performed with the accuracy of $\sim 13\,\%$
  with the center-of-mass energy $500\,{\rm GeV}$ and the integrated
  luminosity $\sim500 \text{--} 1000\,{\rm fb}^{-1}$.

\end{abstract}

\end{titlepage}

\setcounter{page}{1}
\renewcommand{\thefootnote}{\#\arabic{footnote}}
\setcounter{footnote}{0}

\section{Introduction}

It has been known that there exists notable discrepancy between the
experimentally measured and theoretically predicted values of the muon
anomalous magnetic dipole moment ($g-2$).  The Brookhaven E821
experiment \cite{g-2_bnl2010} reported, for $a_\mu=(g-2)/2$,
\begin{equation}
  a_\mu^{\rm (exp)} = (11\,659\,208.9 \pm 6.3) \times 10^{-10}.
\end{equation}
There are several theoretical estimates of the standard-model (SM) value of the muon $g-2$. Based on the analysis of
Refs.~\cite{g-2_hagiwara2011} and \cite{g-2_davier2010} for the hadronic vacuum polarization, the predictions are
\begin{align}
  a_\mu^{\rm (SM)} =
 \begin{cases}
  (11\,659\,182.8 \pm 5.0) \times 10^{-10}, & \mbox{\cite{g-2_hagiwara2011}} \\
  (11\,659\,180.2 \pm 4.9) \times 10^{-10}, & \mbox{\cite{g-2_davier2010}}
 \end{cases} 
  \label{eq:amu_HLMNT}
\end{align}
where we take account of the five-loop QED calculation \cite{Aoyama:2012wk} and the latest update 
of the electroweak contribution \cite{Gnendiger:2013pva}. 
Thus, the difference is estimated as
\begin{align}
  \Delta a_\mu \equiv a_\mu^{\rm (exp)} - a_\mu^{\rm (SM)}
  = 
  \begin{cases}
  (26.1 \pm 8.0) \times 10^{-10}, & \mbox{\cite{g-2_hagiwara2011}} \\
  (28.7 \pm 8.0) \times 10^{-10}. & \mbox{\cite{g-2_davier2010}}
  \end{cases}
  \label{eq:deltaamu}
\end{align}
Hence, there exists more than $3$-$\sigma$ discrepancy between the
experimental and theoretical values.  We call this discrepancy as
``muon $g-2$ anomaly.''  The origin of the muon $g-2$ anomaly is yet
unknown.

If low-energy supersymmetry (SUSY) exists, the SUSY contribution to
the muon $g-2$, denoted as $a_\mu^{\rm (SUSY)}$, can be sizable.  In
particular, when $\tan\beta$, which is a ratio of the vacuum
expectation values of up- and down-type Higgses, is relatively large,
$a_\mu^{\rm (SUSY)}$ can be easily as large as $\Delta a_\mu$
\cite{Lopez:1993vi,Chattopadhyay:1995ae,Moroi:1995yh}.  Thus, it is 
possible that the muon $g-2$ anomaly originates in the SUSY contribution.
The primary purpose of this letter is to point out that we may have a 
chance to test this possibility by reconstructing $a_\mu^{\rm (SUSY)}$, 
if superparticles are found in future collider experiments, and  
if their properties are determined. 

At the leading order, the SUSY contribution to the muon $g-2$ is composed of smuon--neutralino 
and sneutrino--chargino loop diagrams.
In order to reconstruct $a_\mu^{\rm (SUSY)}$, it is necessary to 
understand properties of sleptons, in particular, those of smuons.  
Unfortunately, they may not be well studied at LHC.  On the contrary, 
once the International $e^+e^-$ Linear Collider (ILC) \cite{Behnke:2013xla} 
is built, it is possible to determine them precisely as long as the 
superparticles are within the kinematical reach.

In this letter, we raise a question how and how accurately the SUSY
contribution to the muon $g-2$ can be reconstructed by using ILC
measurements of the parameters of the minimal supersymmetric standard
model (MSSM).  We assume that the muon $g-2$ anomaly is due to the
SUSY contribution.  Since the contribution depends on MSSM parameters,
we concentrate on a particular case where it is dominated by so-called
Bino diagram.  Such a setup is especially interesting, because
sleptons are expected to be within the kinematical reach of ILC
\cite{Endo:2013lva}.  It will be shown that $a_\mu^{\rm (SUSY)}$ can
be reconstructed with the accuracy of $\sim 13\,\%$ for the sample
point we adopt, once ILC runs at the center-of-mass energy
$\sqrt{s}=500\,{\rm GeV}$ and accumulates the integrated luminosity
${\cal L}\sim500 \text{--} 1000\,{\rm fb}^{-1}$.

\section{Framework} \label{sec:framework}
Let us first summarize the framework of the analysis.
The SUSY contribution to the muon $g-2$
strongly depends on MSSM parameters. In this letter, we concentrate on the case
where it is dominated by so-called Bino
 diagram. This situation is realized if the Wino and Higgsino mass
parameters are much larger than the Bino mass parameter.  In this limit, 
the leading contribution is given by (cf.~Ref.~\cite{Moroi:1995yh})
\begin{equation}
  a_\mu^{(\tilde{B})}\equiv
  - \frac{g_{Y}^2}{16\pi^2}
  \frac{ m_\mu M_1 m_{\tilde\mu LR}^2 }
  {m^2_{\tilde{\mu}1}m^2_{\tilde{\mu}2}}
  f^N
  \left(
    \frac{m^2_{\tilde{\mu}1}}{M^2_1},\frac{m^2_{\tilde{\mu}2}}{M^2_1} 
  \right).
  \label{eq:amuBino}
\end{equation}
In the expression, $M_1$ is the Bino mass parameter,
$m_{\tilde{\mu}A}$ ($A=1,2$) is the $A$-th lightest smuon mass, and
$g_Y$ is the gauge coupling constant for U(1)$_Y$, which comes from
the Bino--(s)muons interactions.  Also, $m_{\tilde\mu LR}^2$ is the
left-right mixing parameter in the smuon mass matrix.  The loop
function $f_N$ is defined as
\begin{equation}
  f^N(x,y) = xy 
  \left[ 
    \frac{-3 + x + y + xy}{(x-1)^2(y-1)^2} 
    + \frac{2x \ln x}{(x-y)(x-1)^3} 
    - \frac{2y \ln y}{(x-y)(y-1)^3} 
  \right].
\end{equation}
It is notable that $a_\mu^{(\tilde{B})}$ can be as large as $\Delta
a_\mu$ especially when the Higgsinos are heavy, since $m^2_{\tilde\mu
  LR}$ is enhanced when $\mu\tan\beta$ is large, where $\mu$ is the
Higgsino mass parameter.  In contrast, the other contributions to the
muon $g-2$, including those from the second-lightest or heavier
neutralino, are suppressed if the Higgsinos are decoupled.

The contribution $a_\mu^{ (\tilde B)}$ can be reconstructed if the
Bino mass, smuon masses, and the left-right mixing parameter
$m_{\tilde\mu LR}^2$ are known.  As we will see below, they are
expected to be determined very accurately at ILC, if the sleptons and
the Bino-like neutralino are within the kinematical reach.  In fact,
since the ILC measurements are very precise, the leading approximation
given in Eq.\ \eqref{eq:amuBino} may not be accurate enough to be
compared with the ILC analyses.  In addition, there is a subtlety in
relating the {\em gaugino} coupling constants with the gauge coupling
constants in particular when some of the superparticles are relatively
heavy
\cite{Hikasa:1995bw,Nojiri:1996fp,Cheng:1997sq,Nojiri:1997ma,Endo:2013lva}.
Thus, we will use more complete formula for $a_\mu^{\rm (SUSY)}$.

The full one-loop level formula for $a_\mu^{\rm(SUSY)}$ consists of the contribution from smuon--neutralino loop diagrams and sneutrino--chargino diagrams.
The smuon--neutralino contribution is given by~\cite{Moroi:1995yh}
\begin{equation}
  a_\mu^{(\tilde\chi^0)} =
  \frac{1}{16\pi^2} \sum_{A, X} \frac{m_\mu^2}{m_{\tilde{\mu}A}^2}
  \left[
    -\frac{1}{12} 
    \left[ (N^{\mu_L}_{AX})^2 + (N^{\mu_R}_{AX})^2 \right] F^N_1(x_{AX})
    -\frac{m_{\tilde\chi^0_X}}{3 m_\mu} N^{\mu_L}_{AX} N^{\mu_R}_{AX} F^N_2(x_{AX})
  \right],
  \label{eq:amuNeutralino}
\end{equation}
which includes the leading contribution $a_\mu^{(\tilde B)}$.
Here, $m_{\tilde\chi^0_X}$ ($X=1\text{--}4$) is the neutralino mass,
$x_{AX}=m_{\tilde\chi^0_X}^2/m_{\tilde{\mu}A}^2$, and the loop functions are
\begin{align}
  F^N_1(x) &= \frac{2}{(1-x)^4} \left[ 1-6x+3x^2+2x^3-6x^2\ln x \right],
  \\
  F^N_2(x) &= \frac{3}{(1-x)^3} \left[ 1-x^2+2x\ln x \right].
\end{align}
In addition, $N^{\mu_L}_{AX}$ and $N^{\mu_R}_{AX}$ are
neutralino--muon--smuon coupling constants.  Parameterizing interactions
of neutralinos as
\begin{equation}
  {\cal L}_{\rm int} = 
  \sum_{\ell =e,\mu,\tau}
  \sum_{A,X} 
  \bar{\chi}^0_X (N^{\ell_L}_{AX} P_L + N^{\ell_R}_{AX} P_R) \ell\,
  \tilde{\ell}_A^\dagger
  + {\rm h.c.},
\end{equation}
the coefficients are
\begin{align}
  N^{\ell_L}_{AX} &= 
  \frac{1}{\sqrt{2}} \tilde{g}_{Y,L} (U_{\chi^0})_{X\tilde{B}}
  (U_{\tilde{\ell}})_{AL}
  +\frac{1}{\sqrt{2}} \tilde{g}_{2} (U_{\chi^0})_{X\tilde{W}}
  (U_{\tilde{\ell}})_{AL}
  -y_\ell (U_{\chi^0})_{X\tilde{H}_d} (U_{\tilde{\ell}})_{AR},
  \label{eq:N^L}
  \\
  N^{\ell_R}_{AX} &= 
  -\sqrt{2} \tilde{g}_{Y,R} (U_{\chi^0})_{X\tilde{B}}
  (U_{\tilde{\ell}})_{AR}
  -y_\ell (U_{\chi^0})_{X\tilde{H}_d} (U_{\tilde{\ell}})_{AL}.
  \label{eq:N^R}
\end{align}
Here, $y_\ell$ is the Yukawa coupling constants in the superpotential.
The unitary matrices $U_{\chi^0}$ and $U_{\tilde{\ell}}$ diagonalize
the mass matrices of neutralinos and sleptons, respectively.  It is
assumed that soft SUSY breaking parameters of the sleptons are
independent of the generation, and all the complex phases of the SUSY
parameters are negligibly small, in order to avoid too large
lepton-flavor violations and electric dipole moments.  Then, the
slepton masses are obtained from the slepton mass matrix,
\begin{equation}
  {\cal M}_{\tilde{\ell}}^2 = 
  \begin{pmatrix}
    m_{\tilde{\ell}{LL}}^2
    & 
    m_{\tilde{\ell}{LR}}^2
    \\
    m_{\tilde{\ell}{LR}}^2
    & 
    m_{\tilde{\ell}{RR}}^2
  \end{pmatrix},
  \label{eq:MassMatrix}
\end{equation}
which is diagonalized by the following unitary matrix,
\begin{equation}
  U_{\tilde\ell} = 
  \begin{pmatrix}
  \cos\theta_{\tilde\ell} & \sin\theta_{\tilde\ell} \\
 -\sin\theta_{\tilde\ell} & \cos\theta_{\tilde\ell}
  \end{pmatrix}.
  \label{eq:UnitaryMatrix}
\end{equation}
The slepton mixing angle satisfies the relation,
\begin{equation}
  m_{\tilde\ell{LR}}^2 = 
  \frac{1}{2} (m_{\tilde\ell1}^2 - m_{\tilde\ell2}^2) 
  \sin 2 \theta_{\tilde\ell}.
  \label{eq:sin2thl}
\end{equation}
This relation will play an important role in the following discussion.

It should be noticed that the coupling constants for the
gaugino--lepton--slepton vertices, or the gaugino coupling constants,
deviate from the ordinary gauge coupling constants
\cite{Hikasa:1995bw,Nojiri:1996fp,Cheng:1997sq,Nojiri:1997ma,Endo:2013lva}.
In Eqs.~\eqref{eq:N^L} and \eqref{eq:N^R}, the parameters
$\tilde{g}_{Y,L}$, $\tilde{g}_{Y,R}$, and $\tilde{g}_{2}$ are
introduced to take account of such an effect.  In the SUSY limit,
$\tilde{g}_{Y,L}=\tilde{g}_{Y,R}=g_Y$ and $\tilde{g}_{2}=g_2$ are
satisfied (with $g_Y$ and $g_2$ being the gauge coupling constants of
U(1)$_Y$ and SU(2)$_L$, respectively).  These relations are violated
when some superparticles are (much) heavier than the sleptons.  In the
case where all the superparticles except for sleptons and the Bino are
heavy, we obtain the following approximate formula for
$\tilde{g}_{Y,L}$ and $\tilde{g}_{Y,R}$
(cf.~Ref.~\cite{Endo:2013lva,Fargnoli:2013zda}):
\begin{align}
  \tilde{g}_{Y,L} (Q) &\simeq
  g_Y (Q)
  \left[1 + \frac{1}{4\pi}
    \left(
      4 \alpha_Y \ln \frac{M_{\rm soft}}{Q}
      - \frac{1}{6} \alpha_Y \ln \frac{M_{\tilde H}}{Q}
      + \frac{9}{4} \alpha_2 \ln \frac{M_{\tilde{W}}}{Q}
    \right)
  \right],
  \\
  \tilde{g}_{Y,R} (Q) &\simeq
  g_Y (Q)
  \left[1 + \frac{1}{4\pi}
    \left(
      4 \alpha_Y \ln \frac{M_{\rm soft}}{Q}
      - \frac{1}{6} \alpha_Y \ln \frac{M_{\tilde H}}{Q}
    \right)
  \right],
\end{align}
where $M_{\rm soft}$ is a mass scale of colored superparticles and
heavy Higgses, $M_{\tilde H}$ is the Higgsino mass, $M_{\tilde{W}}$ is the Wino mass, 
and $Q$ $(\sim m_{\tilde\ell})$ is an energy scale.  The differences among 
$g_Y$, $\tilde{g}_{Y,L}$ and $\tilde{g}_{Y,R}$ can be ${\cal O}(1\text{--}10)\%$ 
if $M_{\rm soft}$, $M_{\tilde H}$ and $M_{\tilde{W}}$ are larger than $\sim 1\,{\rm TeV}$. 
Note that the leading contribution of \eqref{eq:amuNeutralino} is proportional to the product 
 $\tilde g_{Y,L}\tilde g_{Y,R}$  (cf.~Eq.~\eqref{eq:amuBino}).
Since the corrections to the gaugino couplings can be sizable,
both of the couplings should be determined directly at ILC.  It is also 
noted that $\tilde{g}_{Y,L}$, $\tilde{g}_{Y,R}$ and $\tilde{g}_{2}$ are universal for 
(at least) light generations.

\begin{table}[t]
\centering
    \caption{Parameters and mass spectrum and at our sample point.  The
      masses are in units of GeV, and $\tilde\ell$ denotes selectrons
      and smuons.}
    \label{table:MSSMparams}
\vspace{0.5em}
    \begin{tabular}{l|ccccccc|c}
      \hline\hline
      {Parameters} & 
      $m_{\tilde{\ell}1}$ & 
      $m_{\tilde{\ell}2}$ & 
      $m_{\tilde{\tau}1}$ & 
      $m_{\tilde{\tau}2}$ & 
      $m_{\tilde{\chi}^0_1}$ &
      $\sin \theta_{\tilde{\mu}}$ &
      $\sin \theta_{\tilde{\tau}}$& 
     $a_\mu^{\rm(ILC)}$\\
      \hline
      {Values} & 
      126 & 200 & 108 & 210 & 90 &
      0.027 & 0.36&$2.6\times10^{-9}$ \\
      \hline\hline
    \end{tabular}
\end{table}

In the following discussion, we choose a specific sample point to make
our discussion concrete and quantitative.  The mass spectrum at the
sample point is summarized in Table \ref{table:MSSMparams}.  All the
sleptons and the lightest neutralino are within the reach of ILC with
$\sqrt{s}=500\,{\rm GeV}$.  Their masses are set to be close to those
of the SPS1a$'$ benchmark point \cite{AguilarSaavedra:2005pw}, so that
results of the previous ILC studies can be applied.  The lighter
sleptons are chosen to be almost left-handed in order to avoid LHC
limits (see below). The lightest neutralino mass is $90\,{\rm GeV}$,
which is the lightest superparticle among the MSSM ones including
sneutrinos.  Other superparticles such as colored ones as well as
Winos and Higgsinos are assumed to be so heavy that they are not
observed at LHC nor ILC (so that their masses are different from those
for SPS1a$'$).\footnote
{ This setup is minimal to reconstruct the SUSY contributions to the
  muon $g-2$. If some of the heavy superparticles such as Winos would be
  additionally discovered, the reconstruction could be improved.  }
Trilinear couplings of sleptons, $A_{\tilde\ell}$, are set to be zero.
The left-right mixing parameter, $m_{\tilde\mu LR}^2$, (or
equivalently $\mu\tan\beta$) is chosen to realize that $a_\mu^{\rm
  (ILC)}$ defined in Eq.~\eqref{eq:amu_ILC} becomes equal to
$2.6\times 10^{-9}$, which is close to the central value of the
current discrepancies \eqref{eq:deltaamu};
$\mu\tan\beta=6.1\times10^3\ {\rm GeV}$.

The mass spectrum is consistent with present collider limits. Light
sleptons decaying to the lightest neutralino are searched for by
studying the di-lepton signatures at LHC
\cite{ATLAS2013049,CMSPASSUS13006}. Our sample point is not excluded
because masses of the left-handed selectron and smuon are close to
that of the neutralino. Also, constraints on the right-handed ones are
weak, since the production cross sections are small. On the other
hand, collider limits on the stau mass is weaker as $m_{\tilde\tau1}
> 81.9\,{\rm GeV}$ at $95\,\%$ CL by LEP \cite{PDG}. Exclusions from
the three-lepton searches at LHC \cite{ATLAS2013035,CMSPASSUS13006}
are also negligible, since Winos and Higgsinos are heavy.

\section{Fun with ILC}

In the rest of this letter, we discuss how and how accurately the 
SUSY contribution to the muon $g-2$ is determined at ILC. 
At the sample point, only the sleptons 
and the lightest neutralino are within the reach of ILC. 
The observed neutralino is identified as Bino-like by absent signals 
of charginos, since neutral Winos or Higgsinos are associated by charged 
partners.  
Let us define the following quantity (cf.~Eq.~\eqref{eq:amuNeutralino}),
\begin{equation}
  a_\mu^{\rm (ILC)} \equiv 
  \frac{1}{16\pi^2} \sum_{A} \frac{m_\mu^2}{m_{\tilde{\mu}A}^2}
  \left[ -\frac{1}{12}
    \left[(\hat{N}^{\mu_L}_{A})^2 + (\hat{N}^{\mu_R}_{A})^2\right] F^N_1(x_{A1})
    - \frac{m_{\tilde\chi^0_1}}{3 m_\mu}
    \hat{N}^{\mu_L}_{A} \hat{N}^{\mu_R}_{A} F^N_2(x_{A1})
  \right],
  \label{eq:amu_ILC}
\end{equation}
which depends only on ILC observables. The parameters are defined as
\begin{align}
  \hat{N}^{\mu_{L}}_{A} &\equiv 
  \left[
    N^{\mu_{L}}_{A1}
  \right]_{(U_{\chi^0})_{1\tilde{H}_d}\rightarrow 0}
  =
  \frac{1}{\sqrt{2}} \tilde{g}_{1,L}^{\rm (eff)}
  (U_{\tilde{\mu}})_{AL},
  \\
  \hat{N}^{\mu_{R}}_{A} &\equiv 
  \left[
    N^{\mu_{R}}_{A1}
  \right]_{(U_{\chi^0})_{1\tilde{H}_d}\rightarrow 0}
  =
  -\sqrt{2} \tilde{g}_{1,R}^{\rm (eff)}
  (U_{\tilde{\mu}})_{AR},
\end{align}
where
\begin{align}
\tilde{g}_{1,L}^{\rm (eff)}&\equiv \tilde{g}_{Y,L}
(U_{\chi^0})_{1\tilde{B}} +\tilde{g}_{2} (U_{\chi^0})_{1\tilde{W}},\\
\tilde{g}_{1,R}^{\rm (eff)}&\equiv
\tilde{g}_{Y,R}(U_{\chi^0})_{1\tilde{B}}.
\end{align}
The smuon mixing angle can be determined 
if the left-right mixing parameter of the smuon, $m_{\tilde\mu LR}^2$, 
as well as the smuon mass eigenvalues are determined, as noticed from 
Eq.~\eqref{eq:sin2thl}. 
Thus, $a_\mu^{\rm (ILC)}$ can be reconstructed if the following quantities 
are known:
\begin{equation}
  m_{\tilde\mu1},\ m_{\tilde\mu2},\ m_{\tilde\mu LR}^2,\ 
  m_{\tilde\chi_1^0},\ 
  \tilde{g}_{1,L}^{\rm (eff)},\
  \tilde{g}_{1,R}^{\rm (eff)}.
  \label{eq:parameters}
\end{equation}
In the following of this section, we consider the reconstruction of $a_\mu^{\rm (ILC)}$
with the determinations of these parameters at ILC.

The full SUSY contribution $a_\mu^{\rm(SUSY)}$ contains the contribution from charginos and heavier neutralinos.
Difference between $a_\mu^{\rm (SUSY)}$ and $a_\mu^{\rm (ILC)}$ will be discussed in
Sec.~\ref{sec:reconstr-amu}.
We show that, at the sample point, future experiments can confirm that 
$a_\mu^{\rm(SUSY)}$ is dominated by $a_\mu^{\rm(ILC)}$.

\subsection{Determination of the left-right mixing}

One of the crucial parameters to calculate $a_\mu^{\rm (ILC)}$ is the left-right mixing 
parameter $m_{\tilde{\mu}LR}^2$. In order to reconstruct the 
smuon unitary matrix $U_{\tilde{\mu}}$, it is necessary to determine 
the mixing angle $\theta_{\tilde{\mu}}$ or $m_{\tilde{\mu}{LR}}^2$.
Although smuons are produced at ILC, it is challenging to
determine them from  smuon measurements.
Importantly, however, $m_{\tilde{\mu}{LR}}^2$ can be obtained from studies of staus.  
The mixing parameters are scaled by the lepton masses as
\begin{equation}
  m_{\tilde{\mu}{LR}}^2
  =
  \frac{m_\mu}{m_\tau} m_{\tilde{\tau}{LR}}^2.
  \label{eq:mlrratio}
\end{equation}
This relation is valid in the limit of
$A_{\tilde{\ell}}\ll\mu\tan\beta$, where $A_{\tilde{\ell}}$ is the
trilinear coupling constant of the slepton $\tilde{\ell}$ normalized
by the corresponding Yukawa coupling constant. This is the case at our
sample point.  Using Eq.~\eqref{eq:sin2thl}, $m_{\tilde{\tau}{LR}}^2$
is determined if $\sin 2\theta_{\tilde{\tau}}$ as well as the mass
eigenvalues, $m_{\tilde{\tau}1}$ and $m_{\tilde{\tau}2}$, are
measured.  Its accuracy is estimated as
\begin{equation}
\left( \delta m_{\tilde\tau{LR}}^2 \right)^2 = 
\left( \frac{\partial m_{\tilde\tau{LR}}^2}{\partial m_{\tilde\tau1}} \right)^2 
\left( \delta m_{\tilde\tau1} \right)^2 + 
\left( \frac{\partial m_{\tilde\tau{LR}}^2}{\partial m_{\tilde\tau2}} \right)^2 
\left( \delta m_{\tilde\tau2} \right)^2 + 
\left( \frac{\partial m_{\tilde\tau{LR}}^2}{\partial \sin 2\theta_{\tilde\tau}} \right)^2 
\left( \delta \sin 2\theta_{\tilde\tau} \right)^2,
\label{eq:delta_mLR}
\end{equation}
where the derivatives are evaluated at the sample point. In particular, 
$\sin 2\theta_{\tilde{\tau}}$ can be naturally as large as
$O(0.1)$ in the parameter region where $a_\mu^{\rm (SUSY)}\simeq
\Delta a_\mu$.

First of all, the stau mass eigenvalues, $m_{\tilde{\tau}1}$ and
$m_{\tilde{\tau}2}$, can be determined by measuring the endpoints of
the energy distribution of $\tau$ decay products from the stau decay,
$\tilde{\tau}^\pm\rightarrow\tau^\pm\tilde\chi_0^1$
\cite{Nojiri:1996fp}.  For such an analysis, information about the
lightest neutralino mass is also needed; measurement of
$m_{\tilde{\chi}^0_1}$ will be discussed in the next subsection.

The measurement of stau masses at ILC is
discussed in detail in Ref.~\cite{Bechtle:2009em}.  It is claimed that
the mass can be determined with the accuracy of $\sim 0.1\,\%$
($3\,\%$) for lighter (heavier) stau with $\sqrt{s}=500\,{\rm GeV}$,
$(P_{e+},P_{e-})=(-0.3,+0.8)$ and the integrated luminosity ${\cal
  L}=500\,{\rm fb}^{-1}$.  Here, $P_{e-}$ $(P_{e+})$ is the degree of
transverse polarization of the electron (positron) beam. The right-
(left-) handed polarization corresponds to $P_{e}=+1$ $(-1)$.  The
analysis depends on details of the mass spectrum.  In
Ref.~\cite{Bechtle:2009em}, the SPS1a$'$ benchmark point is adopted,
and signal regions are optimized for it. In particular, lighter
(heavier) stau is almost right-handed (left-handed), and the
neutralino mass is $98\,{\rm GeV}$. These are different from our
sample point and could affect the accuracy. In fact, the energy
profile of the decay products of $\tau$ depends on the helicity of
$\tau$ \cite{Tsai:1971vv}.  For instance, jet energy from $\tau \to
\pi \nu$ is likely to be harder for $\tau_R$ compared to $\tau_L$
\cite{Nojiri:1994it}.  Also, with the polarization used in
Ref.~\cite{Bechtle:2009em}, the production cross section of the
lighter (heavier) stau at our model point is smaller (larger) than
those at SPS1a$'$.  On the other hand, the endpoint energies of
$\tau$-jet increase, as $m_{\tilde\chi^0_1}$ decreases
(cf.~Ref.~\cite{Nojiri:1994it}). Then, the contamination of the
background due to the process $\gamma\gamma\rightarrow\tau^+\tau^-$ is
reduced \cite{Bechtle:2009em}.  In this letter, we simply adopt the
accuracy of $0.1\,\%$ and $3\,\%$ as our canonical values for the mass
measurements of the staus.\footnote
{\label{footnote:stau_mass} Dedicated studies of the threshold
  production of $\tilde{\tau}_2$ can improve the accuracy of its mass
  measurement \cite{Grannis:2002xd,Baer:2013cma}. At the Snowmass SM2
  benchmark point, which is close to the SPS1a point, $\delta
  m_{\tilde{\tau}2} \sim 1\,{\rm GeV}$ is available for
  $m_{\tilde{\tau}2} = 206\,{\rm GeV}$, where $\sqrt{s}=500\,{\rm
    GeV}$ and ${\cal L}=1000\,{\rm fb}^{-1}$ with the electron
  polarization of $80\,\%$, while no polarization for the positron.  }

\begin{figure}[tbp]
 \begin{center}
 \includegraphics[width=7cm]{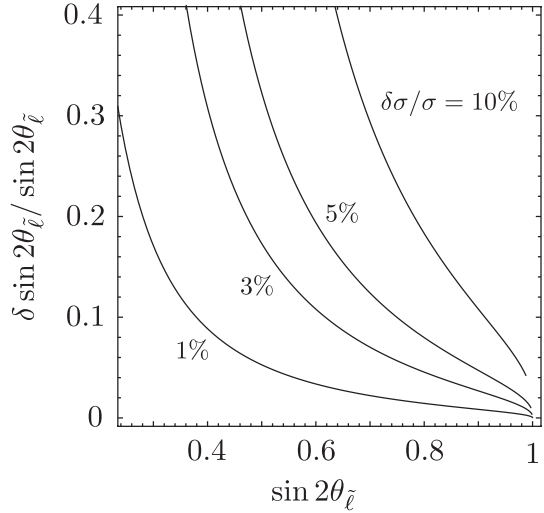}
 \end{center}
 \caption{Accuracies of the determination of $\sin 2\theta_{\tilde{\tau}}$ from the measurement of the cross section $\sigma(e^+e^-\to\tilde\tau_1\tilde\tau_1)$ as a function of the stau mixing angle with the accuracy of the cross section determination of $10\,\%$, $5\,\%$, $3\,\%$ and $1\,\%$ from top to bottom. The mass measurement is assumed to be sufficiently precise.
}
\label{fig:angle_measurement}
\end{figure}

Next, the mixing angle $\theta_{\tilde{\tau}}$ can be determined from the
measurements of the cross sections of stau pair production processes.\footnote
{Alternatively, the stau mixing angle can be determined by measuring the $\tau$ 
polarization from energy profile of its decay products \cite{Nojiri:1996fp}. 
However, the cross section measurement provides a better resolution \cite{Bechtle:2009em}.
}
The cross sections are given by \cite{Boos:2003vf}
\begin{equation}
\begin{split}
\sigma(e^+e^-\to\tilde\tau_i\tilde\tau_j) &= \frac{8\pi\alpha^2}{3s}
v^3 \bigg[
c_{ij}^2 \frac{\Delta_Z^2}{\sin^42\theta_W}(\mathcal{P}_{-+}L^2 + \mathcal{P}_{+-}R^2)
\\&\qquad
+\delta_{ij} \frac{1}{16} (\mathcal{P}_{-+} + \mathcal{P}_{+-})
+\delta_{ij}c_{ij}\frac{\Delta_Z}{2\sin^22\theta_W}(\mathcal{P}_{-+}L + \mathcal{P}_{+-}R)
\bigg],
\end{split}
\label{eq:CrossSection}
\end{equation}
where the parameters are defined as 
\begin{align}
&v^2 = [1-(m_{\tilde\tau_i}+m_{\tilde\tau_j})^2/s][1-(m_{\tilde\tau_i}-m_{\tilde\tau_j})^2/s], \\
&\Delta_Z = s/(s-m_Z^2), \\
&c_{11/22} = \frac{1}{2} \left[L+R \pm (L-R) \cos 2\theta_{\tilde{\tau}} \right],\\
&c_{12} = c_{21} = \frac{1}{2}(L-R)\sin 2\theta_{\tilde{\tau}},\\
&L=-\frac{1}{2}+\sin^2\theta_W,\\
&R=\sin^2\theta_W.
\end{align}
The beam polarizations are parameterized as $\mathcal{P}_{\mp\pm} = (1
\mp P_{e-}) (1 \pm P_{e+})$.

We consider productions of lighter staus to determine
$\theta_{\tilde{\tau}}$.  The cross section, $\sigma(\tilde\tau_1) =
\sigma(e^+e^-\to\tilde\tau_1^+\tilde\tau_1^-)$, depends on
$m_{\tilde\tau1}$ and $\theta_{\tilde\tau}$. The accuracy of the
measurement of the stau mixing angle is estimated as
\begin{equation}
\left( \delta \sin 2\theta_{\tilde{\tau}} \right)^2 = 
\left( \frac{\partial \sin 2\theta_{\tilde{\tau}}}{\partial \sigma(\tilde\tau_1)} \right)^2 
\left( \delta \sigma(\tilde\tau_1) \right)^2 +
\left( \frac{\partial \sin 2\theta_{\tilde{\tau}}}{\partial m_{\tilde\tau1}} \right)^2 
\left( \delta m_{\tilde\tau1} \right)^2. 
\label{eq:error1}
\end{equation}
In the sample point, the error is dominated by that of the cross section. 
The stau mass contributes to the cross section only through $v$, and
$m_{\tilde\tau1}$ is (much) smaller than $\sqrt{s}$. Further, mass
of $\tilde{\tau}_1$ can be precisely measured, as mentioned above.  
According to Ref.~\cite{Bechtle:2009em}, the cross section for 
$e^+e^-\rightarrow \tilde{\tau}^+_1\tilde{\tau}^-_1$ can be measured with 
the accuracy of $3.1\,\%$ for SPS1a$'$. Here, the uncertainty originates in the signal
statistics and SUSY background, while those of the luminosity and
efficiencies are assumed to be negligible.  In our sample point, the
production cross section is $\sigma(\tilde\tau_1)=54\,{\rm fb}$ with
$\sqrt{s}=500\,{\rm GeV}$ and $(P_{e+},P_{e-})=(-0.3,+0.8)$, which
is smaller than $\sigma(\tilde\tau_1)=135\,{\rm fb}$ at SPS1a$'$.  By
supposing the same acceptance as Ref.~\cite{Bechtle:2009em}, the
statistical uncertainty increases from $1/\sqrt{N_{\rm sig}}|_{\rm
  SPS1a'} = 2.1\,\%$ to $3.4\,\%$ with ${\cal L}=500\,{\rm fb}^{-1}$,
where $N_{\rm sig}$ is the number of the signals accepted by selections.
On the other hand, our setup is free from the SUSY background, since only sleptons and the lightest neutralino are produced at our sample point. Thus, the accuracy 
of the cross section measurement is estimated to be
$\delta\sigma(\tilde\tau_1)/\sigma(\tilde\tau_1)=3.4\,\%$.\footnote
{
The signal region can be optimized for our sample point.
Since there is no SUSY background, the acceptance could be enhanced and the uncertainty would be reduced.
}

In Fig.~\ref{fig:angle_measurement}, the accuracy of the measurement of
$\sin 2\theta_{\tilde{\tau}}$ is shown.
The accuracy is sensitive to the mixing angle. It becomes better when the
angle approaches to maximal, $\theta_{\tilde{\tau}} = \pi/4$. This is
because $\sigma(\tilde\tau_1)$ depends on $\theta_{\tilde{\tau}}$ via
$\cos 2\theta_{\tilde{\tau}}$. In the sample point, where 
$\sin 2\theta_{\tilde{\tau}}=0.67$, it is expected that $\sin 2\theta_{\tilde{\tau}}$ can be determined 
with the accuracy of $9\,\%$ by applying
$\delta\sigma(\tilde\tau_1)/\sigma(\tilde\tau_1)=3.4\,\%$ (and $\delta
m_{\tilde\tau1}/m_{\tilde\tau1} \sim 0.1\,\%$).

Finally, by combining the uncertainties of the determinations of 
$m_{\tilde{\tau}1}$, $m_{\tilde{\tau}2}$ and $\sin
2\theta_{\tilde{\tau}}$, 
the accuracy of the $m_{\tilde\tau{LR}}^2$ determination 
is estimated by Eq.~\eqref{eq:delta_mLR}.
The uncertainty due to the measurement of the lighter stau mass is
negligible, since $m_{\tilde\tau1}$ and $m_{\tilde\tau2}$ contribute
to $m_{\tilde\tau{LR}}^2$ in the combination of $(m_{\tilde\tau1}^2
- m_{\tilde\tau2}^2)$, and the uncertainty of the heavier stau mass
is larger than that of the lighter one.  Also, correlation of
the errors, $\delta m_{\tilde\tau1}$ and $\delta \sin
2\theta_{\tilde\tau}$, is negligible, since $\delta m_{\tilde\tau1}$
barely affects $\delta \sin 2\theta_{\tilde\tau}$ when $\delta
m_{\tilde\tau1}$ is sufficiently small. As a result, we obtain
$\delta m_{\tilde{\tau}{LR}}^2/m_{\tilde{\tau}{LR}}^2 = 12\,\%$ with
$\delta m_{\tilde\tau2}/m_{\tilde\tau2} = 3\,\%$ and $\delta\sin
2\theta_{\tilde{\tau}}/\sin 2\theta_{\tilde{\tau}} = 9\,\%$. From the
relation \eqref{eq:mlrratio}, $m_{\tilde{\mu}{LR}}^2$ is determined
with the same accuracy, 
\begin{equation}
  \delta m_{\tilde{\mu}{LR}}^2/m_{\tilde{\mu}{LR}}^2 = 12\,\%,
  \label{eq:m2LR}
\end{equation}
in the sample point, where $m_{\tilde{\mu}{LR}}^2 = -645\,{\rm GeV}^2$.

The sign of $\sin 2\theta_{\tilde{\tau}}$ is not determined by the
cross section measurements. It corresponds to the sign of $\mu\tan\beta$.
Consequently, the reconstruction of the SUSY contribution to the muon $g-2$ is possible with two fold ambiguity.
We take the sign of $\mu\tan\beta$ so that the muon $g-2$ anomaly is solved.

There are several comments in order.  (i) In the above analysis, we 
considered the process $e^+e^-\rightarrow\tilde\tau_1\tilde\tau_1$. 
The determination of the slepton mixing angle is possibly improved
if the production cross section of a pair of
$\tilde\ell_1$ and $\tilde\ell_2$ is measured accurately
\cite{Bechtle:2009em}, since it is proportional to $\sin^2
2\theta_{\tilde{\ell}}$. In the sample point, the cross section for
the process $e^+e^-\rightarrow\tilde\tau_1\tilde\tau_2$ becomes
$2.7\,{\rm fb}$ ($3.6\,{\rm fb}$) with $\sqrt{s}=500\,{\rm GeV}$ and
$(P_{e+},P_{e-})=(-0.3,+0.8)$ ($(P_{e+},P_{e-})=(0.3,-0.8)$).
However, we could not find studies about such a process.  In
particular, the acceptance of the signal events as well as the
accuracy of the cross section measurement has not been known.  Thus,
in the present study, we do not use this process.  (ii) The smuon
mixing angle is measured directly in principle from the smuon
production $e^+e^-\rightarrow\tilde\mu^+\tilde\mu^-$.  This measurement 
is possible only when the smuon mixing is sufficiently large. 
It can be maximal when $\tilde\mu_L$ and $\tilde\mu_R$
are almost degenerate in mass (see Ref.~\cite{Endo:2013lva} for
example), whereas it is tiny in our sample point.  (iii) The
(approximate) chiralities of lighter and heavier smuons are fixed by
the sign of $\cos2\theta_{\tilde\mu}$, which can be determined by
measuring the smuon production cross sections.  (iv)
Eq.~\eqref{eq:mlrratio} can be violated if $A_{\tilde \ell}$ depends on
generations.\footnote
{
It is difficult to determine $A_{\tilde\tau}$ and $\mu\tan\beta$
individually in 
$m_{\tilde{\tau}{LR}}^2$ by the stau decays. In fact, it is possible 
if the Higgsinos are light \cite{Boos:2003vf}. However, they are 
decoupled in our sample point. Alternatively, $\tan\beta$ is determined 
if the sneutrino mass is measured precisely, for instance, through the 
decay channel $\tilde\nu \to \tilde\chi^\pm_1 \ell^\mp$ 
(see Ref.~\cite{AguilarSaavedra:2001rg}).
In our sample point, it is difficult to identify the sneutrinos, because 
they decay only to the lightest neutralino. 
}
If they are comparable to the slepton masses, the violation is negligible 
compared to the accuracy of the measurement of $m_{\tilde\mu LR}^2$ at ILC. 
Let us suppose that $A_{\tilde\mu}$ differs by $100\,{\rm GeV}$ from $A_{\tilde\tau}$ 
in $m_{\tilde\ell{LR}}^2=-m_\ell(\mu\tan\beta-A_{\tilde\ell})$. In the sample
point, $m_{\tilde\mu LR}^2$ is mis-measured by $\sim 2\,\%$ if it is
determined by Eq.~\eqref{eq:mlrratio}. This is smaller than the
above ILC uncertainty.%
\footnote{%
   The relations between the lepton masses and the Yukawa coupling constants
   are affected by SUSY radiative correction.
   The correction violates the relation Eq.~\eqref{eq:mlrratio}
   if the slepton soft masses depend on the generation.
   The violation is typically small.
}

\subsection{Mass determinations}

Next, let us consider measurements of $m_{\tilde{\mu}1}$,
$m_{\tilde{\mu}2}$ and $m_{\tilde\chi^0_1}$. If smuon masses are
within the reach of ILC, they can be obtained from productions of the
smuons that decay into neutralinos. The energy spectra of the muons
produced by the smuon decay and  the production threshold are
sensitive to the masses \cite{Tsukamoto:1993gt}.  In
Refs.~\cite{Martyn:2004ew,Freitas:2004re,Baer:2013cma}, the accuracies
are estimated to be $\delta m_{\tilde{\mu}R}=170\,{\rm MeV}$ and
$\delta m_{\tilde\chi^0_1}=210\,{\rm MeV}$ at the SPS1a benchmark
point \cite{Allanach:2002nj}.\footnote
{
The neutralino mass can also be measured from the endpoints in the stau productions. 
However, the resolution is worse \cite{Bechtle:2009em}.
}
Here, the masses are $m_{\tilde{\mu}R}=143\,{\rm GeV}$ and
$m_{\tilde\chi^0_1}=96\,{\rm GeV}$ with ${\rm Br}(\tilde{\mu}_R^\pm
\to \mu^\pm\tilde\chi^0_1)=100\,\%$. The analysis is based on
$\sqrt{s}=400\,{\rm GeV}$, $(P_{e+},P_{e-})=(-0.6,+0.8)$ and ${\cal
  L}=200\,{\rm fb}^{-1}$. Another study of the threshold scans yields
$\delta m_{\tilde{\mu}R}=200\,{\rm MeV}$ for
$m_{\tilde{\mu}R}=135\,{\rm GeV}$ by assuming $10\,{\rm fb}^{-1}$ per
each data point with $(P_{e+},P_{e-})=(+0.3,-0.8)$
\cite{Brau:2012hv,Baer:2013cma}. The uncertainties are statistically
limited. The muon energy spectrum is independent of the smuon
chirality, and the mass resolution is less dependent on the
smuon-neutralino mass splitting \cite{Martyn:2004jc}. Since the
accuracy is limited by signal statistics, we expect $\tilde{\mu}_1$
has a better mass resolution in our sample point. At SPS1a, the
production cross section is $\sigma(\tilde\mu_1)=134\,{\rm fb}$ with
$\sqrt{s}=400\,{\rm GeV}$, and $(P_{e+},P_{e-})=(-0.6,+0.8)$. In our
sample point, it becomes $\sigma(\tilde\mu_1)=154\,{\rm fb}$ with
$\sqrt{s}=500\,{\rm GeV}$, and $(P_{e+},P_{e-})=(+0.3,-0.8)$. 

The mass measurement of the heavier smuon is studied in detail at
SPS1a$'$ by Ref.~\cite{Berggren:2009uj}. Here, the heavier smuon is
almost left-handed, and $\sqrt{s}=500\,{\rm GeV}$,
$(P_{e+},P_{e-})=(+0.6,-0.8)$ and ${\cal L}=500\,{\rm fb}^{-1}$ are
used. The resolution can be $\delta m_{\tilde{\mu}L}=100\,{\rm MeV}$
for $m_{\tilde{\mu}L}=190\,{\rm GeV}$ and $m_{\tilde\chi^0_1}=98\,{\rm
  GeV}$ by studying the endpoints.  At SPS1a$'$, most of the produced
$\tilde\mu_L$'s decay into the lightest neutralino and a muon. In our
sample point, all $\tilde\mu_2$ decay into the lightest neutralino and
a muon. The production cross section is $\sigma(\tilde\mu_2)=80\,{\rm
  fb}$ at SPS1a$'$ for $\sqrt{s}=500\,{\rm GeV}$ and
$(P_{e+},P_{e-})=(+0.6,-0.8)$, while it is
$\sigma(\tilde\mu_2)=44\,{\rm fb}$ in our sample point with
$\sqrt{s}=500\,{\rm GeV}$ and $(P_{e+},P_{e-})=(-0.3,+0.8)$. Thus, the
statistical uncertainty is degraded by a factor $1.3$. On the
contrary, the above resolutions could be improved in our sample point,
because SUSY background, for instance, from heavier neutralino
productions, is suppressed. Finally, the accuracy of the neutralino
mass measurement becomes better if studies about the selectron
production processes are combined. In Ref.~\cite{Martyn:2004jc}, it is
claimed that $\delta m_{\tilde\chi^0_1}=80\,{\rm MeV}$ is achieved at
SPS1a.  

In the present analysis, we simply assume
\begin{equation}
  \delta m_{\tilde{\mu}1}=200\,{\rm MeV},~~~
  \delta m_{\tilde{\mu}2}=200\,{\rm MeV},~~~
  \delta m_{\tilde\chi^0_1}=100\,{\rm MeV},
\end{equation}
at the sample point. Then, in the reconstruction of
$a_\mu^{\rm (ILC)}$, the uncertainties in the mass measurements of
smuons and neutralino are less important than that of
$m_{\tilde{\mu}{LR}}^2$.

\subsection{Coupling measurements}

The coupling constants $\tilde{g}_{1,L}^{\rm (eff)}$ and
$\tilde{g}_{1,R}^{\rm (eff)}$ are hardly determined directly from the
smuon production processes.  Instead, they are available from
selectron productions \cite{Nojiri:1996fp,Cheng:1997sq}, because they
are common in light generations.  Since the Yukawa coupling constant
of the electron is negligibly small,
$(U_{\tilde{e}})_{1L}=(U_{\tilde{e}})_{2R}=1$ holds with very high
accuracy.  (Thus, we call lighter and heavier selectrons as
$\tilde{e}_L$ and $\tilde{e}_R$, respectively.)  Consequently, we
obtain
\begin{equation}
  N^{e_L}_{11} = 
  \frac{1}{\sqrt{2}} \tilde{g}_{1,L}^{\rm (eff)},~~~
  {N}^{e_R}_{21} = 
  -\sqrt{2} \tilde{g}_{1,R}^{\rm (eff)}.
\end{equation}
 Cross sections for the selectron production processes 
depend on $N^{e_L}_{11}$ and ${N}^{e_R}_{21}$ through the $t$-channel 
neutralino-exchange diagrams. Thus, $\tilde{g}_{1,L}^{\rm (eff)}$ and 
$\tilde{g}_{1,R}^{\rm (eff)}$ can be measured by studying the selectron 
production cross sections as long as contributions of heavier neutralinos 
are known.

In Refs.~\cite{Freitas:2002gh,Freitas:2003yp,Kilian:2006he}, it is claimed that the
Bino coupling with the (s)electrons can be determined with the
accuracy of $0.18\,\%$ from the measurements of the production cross
section of $\tilde e_R^+\tilde e_R^-$. Here, the beam configuration is
$\sqrt{s}=500\,{\rm GeV}$ with ${\cal L}=500\,{\rm fb}^{-1}$ and the
polarizations of $80\,\%$ (electron) and $50\,\%$ (positron).  In the
analysis, the SPS1a benchmark point is adopted, in which the selectron
mass is $m_{\tilde eR}=143\,{\rm GeV}$. Here, all the neutralino
masses are assumed to be measured by their productions at ILC.  The
production cross section of $\tilde e_R^+\tilde e_R^-$ is very
sensitive to $\tilde{g}_{1,R}^{\rm (eff)}$.  It can be estimated that
the accuracy of the measurement of the $\tilde e_R^+\tilde e_R^-$
cross section should be better than $0.9\,\%$ to determine the
coupling at the $0.18\,\%$ level.  We reinterpret the result of
Refs.~\cite{Freitas:2002gh,Freitas:2003yp} to estimate how accurately
$\tilde{g}_{1,R}^{\rm (eff)}$ can be measured in the sample point.
Let us assume that the accuracy of the cross section measurement is
limited by the signal statistics, and that the acceptance at our
sample point is the same as that in SPS1a. We estimate that the
precision of Refs.~\cite{Freitas:2002gh,Freitas:2003yp} is simply
scaled by $\sqrt{N_{\rm sig}}$.  At SPS1a, the cross section is
$\sigma(\tilde e_R^+\tilde e_R^-)=809\,{\rm fb}$ for
$\sqrt{s}=500\,{\rm GeV}$ and $(P_{e+},P_{e-})=(-0.5,+0.8)$, while
$\sigma(\tilde e_R^+\tilde e_R^-)=316\,{\rm fb}$ in our sample point
for $\sqrt{s}=500\,{\rm GeV}$ and $(P_{e+},P_{e-})=(-0.3,+0.8)$ with
assuming that Winos and Higgsinos are decoupled.  Then, the
experimental uncertainty of the cross section measurement is degraded
to be about $1.5\,\%$.  
We emphasize that Winos and Higgsinos
are assumed to be undiscovered in our sample point. In addition to the 
lightest neutralino, heavier neutralinos, which are mostly composed of 
Winos and Higgsinos, may be exchanged in the $t$-channel diagrams, and 
contribute to the selectron production cross sections. 
In the process $e^+e^- \to \tilde e_R^+\tilde
e_R^-$, their contamination to the gaugino coupling constant
measurement is very small, because they appear only through the mixing
between the Bino and the Higgsinos. The direct interactions of the
Higgsinos to the (s)electron are negligible due to a tiny coupling. In
the case when the Higgsinos are heavier than $500\,{\rm GeV}$
($1\,{\rm TeV}$), we estimate that $\tilde{g}_{1,R}^{\rm (eff)}$
involves a theoretical uncertainty of $0.4\,\%$ ($0.1\,\%$). As a
result, the coupling is expected to be determined with the accuracy of
about $0.7\,\%$ ($0.4\,\%$) in total. Hereafter, we adopt a slightly
conservative value,
\begin{equation}
  \delta\tilde{g}_{1,R}^{\rm (eff)}/\tilde{g}_{1,R}^{\rm (eff)}=1\,\%.
\end{equation}
This uncertainty is sub-dominant in the reconstruction of
$a_\mu^{\rm (ILC)}$ compared to that in $m_{\tilde{\mu}{LR}}^2$.

The gaugino coupling to the left-handed (s)electron is measured from
the production cross section of the left-handed selectrons. In
particular, those of the processes, $e^+e^- \to
\tilde{e}_R^+\tilde{e}_L^-$ or $\tilde{e}_L^+\tilde{e}_R^-$, are
sensitive to $\tilde{g}_{1,L}^{\rm (eff)}$ (as well as
$\tilde{g}_{1,R}^{\rm (eff)}$).\footnote
{ The process, $e^+e^- \to \tilde{e}_L^+\tilde{e}_L^-$, also involves
  $\tilde{g}_{1,L}^{\rm (eff)}$. However, its cross section depends on
  the Wino coupling $\tilde{g}_2$ as well as $\tilde{g}_{Y,L}$ mainly
  through the $t$-channel Wino exchange diagram.}
The cross section can be measured precisely at ILC
\cite{Tsukamoto:1993gt}. In Ref.~\cite{Weiglein:2004hn}, its accuracy 
is claimed to be $\sim 2\,\%$ for $m_{\tilde eR}=143\,{\rm GeV}$ and 
$m_{\tilde eL}=202\,{\rm GeV}$. Here, the SPS1a point is adopted with 
$\sqrt{s}=500\,{\rm GeV}$ and $(P_{e+},P_{e-})=(-0.6,-0.8)$, though
the luminosity is not explicitly shown.
The selectron production processes are
discriminated from each others by the electron energy and by changing
the beam polarization especially of the positron
\cite{Tsukamoto:1993gt,Martyn:2004ew,MoortgatPick:2005cw}.\footnote{%
The heavier selectron 
may be identified by its decay products, if it has sizable branching ratio,
for instance, of $\tilde{e} \to e \tilde\chi^0_2 (\to
\tau^+\tau^-\tilde\chi^1_0)$ \cite{Freitas:2003yp}. However, both of
the selectrons decay directly into the lightest neutralino in our
sample point.}
In fact,
the analysis in Ref.~\cite{MoortgatPick:2005cw} shows that the neutralino 
coupling can be measured at similar accuracy as those
in Ref.~\cite{Freitas:2002gh,Freitas:2003yp} by changing the
polarization. Unfortunately, the acceptance
as well as the accuracy of the cross section measurement is not found
in the literature. In this letter, we assume that
$\sigma(\tilde{e}_L^+\tilde{e}_R^-)$ is measured with the accuracy of 
a few percents. Numerically, if it is determined at the $2\,\%$ ($4\,\%$)
level, the accuracy of $\tilde{g}_{1,L}^{\rm (eff)}$ is estimated to be
about $1\,\%$ ($2\,\%$), where  
$\delta\tilde{g}_{1,R}^{\rm (eff)}/\tilde{g}_{1,R}^{\rm (eff)}=1\,\%$ is 
applied. 

In addition to the experimental uncertainty, the process $e^+e^- \to
\tilde{e}_L^+\tilde{e}_R^-$ involves the $t$-channel exchange diagrams 
of heavier neutralinos. They contribute to the cross section via the 
Wino--Higgsino and Bino--Higgsino mixings. Their contamination to the 
measurement of $\tilde{g}_{1,L}^{\rm (eff)}$ depends on their masses. 
Assuming that Wino and Higgsino masses are above $500\,{\rm GeV}$, 
we estimate that $\tilde{g}_{1,L}^{\rm (eff)}$ involves a theoretical 
(systematic) uncertainty of $0.9\,\%$, while it is reduced to be 
$0.2\,\%$ for $M_{\tilde{W},\tilde{H}}>1\,{\rm TeV}$. On the other hand,
contaminations from corrections to the Wino coupling with the
(s)electrons are smaller than it. As a result, the accuracy of the
measurement of the gaugino coupling is estimated to be
\begin{equation}
  \delta\tilde{g}_{1,L}^{\rm (eff)}/\tilde{g}_{1,L}^{\rm (eff)}=
  \text{a few}\,\%\,({\rm exp})+1\,\%\,({\rm th}),
\end{equation}
or better. Here, the first term in the right-hand side comes from the 
measurement of the cross section for $e^+e^- \to \tilde{e}_L^+\tilde{e}_R^-$, 
and the second term is due to the contamination from the undiscovered 
Winos and Higgsinos. Then, the uncertainty is sub-dominant in the reconstruction 
of $a_\mu^{\rm (ILC)}$ compared to that in $m_{\tilde{\mu}{LR}}^2$.

\begin{table}[t]
  \centering
    \caption{Observables necessary for the reconstruction of
      $a_\mu^{\rm (ILC)}$, and their uncertainties
      with $\sqrt{s}=500$ GeV and ${\cal L}\sim500\text{--}1000\,\text{fb}^{-1}$. 
      Processes relevant to determine each observable are
      also shown. The second and third rows are the information 
      to determine $m_{\tilde{\mu}{LR}}^2$.
      For the determination of $m_{\tilde\chi^0_1}$, analyses 
      of the productions of selectrons and smuons are combined.  
      The uncertainties in $\tilde{g}_{1,L}^{\rm (eff)}$ are 
      those from the experiment and theory, respectively.}
    \label{table:accuracy}
\vspace{0.5em}
    \begin{tabular}{lllll}
      \hline\hline
      {$X$} & {$\delta X$} & {$\delta_X a_\mu^{\rm (ILC)}$} & {Process} \\
      \hline
      $m_{\tilde{\mu}{LR}}^2$ & $12\,\%$ & $13\,\%$ 
      & $e^+e^- \rightarrow \tilde{\tau}^+\tilde{\tau}^-$
      & (cross section, endpoint)
      \\
      ($\sin 2\theta_{\tilde{\tau}}$) & ($9\,\%$) & $-$
      & $e^+e^- \rightarrow \tilde{\tau}^+_1\tilde{\tau}^-_1$
      & (cross section)
      \\
      ($m_{\tilde\tau2}$) & ($3\,\%$) & $-$
      & $e^+e^- \rightarrow \tilde{\tau}^+_2\tilde{\tau}^-_2$
      & (endpoint)
      \\
      $m_{\tilde{\mu}1}$, $m_{\tilde{\mu}2}$ & $200\,{\rm MeV}$ & $0.3\,\%$ 
      & $e^+e^- \rightarrow \tilde{\mu}^+\tilde{\mu}^-$
      & (endpoint)
      \\
      $m_{\tilde\chi^0_1}$ & $100\,{\rm MeV}$ & $< 0.1\,\%$ 
      & $e^+e^- \rightarrow \tilde{\mu}^+\tilde{\mu}^-/\tilde{e}^+\tilde{e}^-$
      & (endpoint)
      \\
      $\tilde{g}_{1,L}^{\rm (eff)}$ & a few$+1\,\%$  & a few$+1\,\%$ 
      & $e^+e^- \rightarrow \tilde{e}_L^+\tilde{e}_R^-$
      & (cross section)
      \\
      $\tilde{g}_{1,R}^{\rm (eff)}$ & $1\,\%$  & $0.9\,\%$ 
      & $e^+e^- \rightarrow \tilde{e}_R^+\tilde{e}_R^-$
      & (cross section)
      \\
      \hline\hline
    \end{tabular}
\end{table}

\subsection{Reconstruction of the SUSY contribution to muon $g-2$}\label{sec:reconstr-amu}

Now let us discuss the accuracy of the
reconstruction of $a_\mu^{\rm (ILC)}$ with ILC.  The accuracy is
estimated by summing all the errors induced by these parameters in
quadrature as
\begin{align}
  \delta a_\mu^{\rm (ILC)} \equiv
    \sqrt{ \sum_X \left( \delta_X a_\mu^{\rm (ILC)} \right)^2 },~~~
  \delta_X a_\mu^{\rm (ILC)} \equiv 
    \frac{\partial a_\mu^{\rm (ILC)}}{\partial X} \delta X ,
  \label{eq:damuB}
\end{align}
where $X=m_{\tilde{\mu}{LR}}^2$, $m_{\tilde{\mu}1}$,
$m_{\tilde{\mu}2}$, $m_{\tilde\chi^0_1}$, $\tilde{g}_{1,L}^{\rm
  (eff)}$, and $\tilde{g}_{1,R}^{\rm (eff)}$. 
In Table \ref{table:accuracy}, their uncertainties are summarized.
Consequently, we estimate $\delta a_\mu^{\rm (ILC)}$ as
\begin{align}
  \delta a_\mu^{\rm (ILC)}/a_\mu^{\rm (ILC)} = 13\,\%,
  \label{eq:accracy_gmin2}
\end{align}
taking $\delta\tilde{g}_{1,L}^{\rm (eff)}\leq3\,\%$.  The dominant
error originates in the determination of the left-right mixing
parameter $m_{\tilde\mu LR}^2$.

\begin{figure}[t]
 \centering
    \includegraphics[width=7cm]{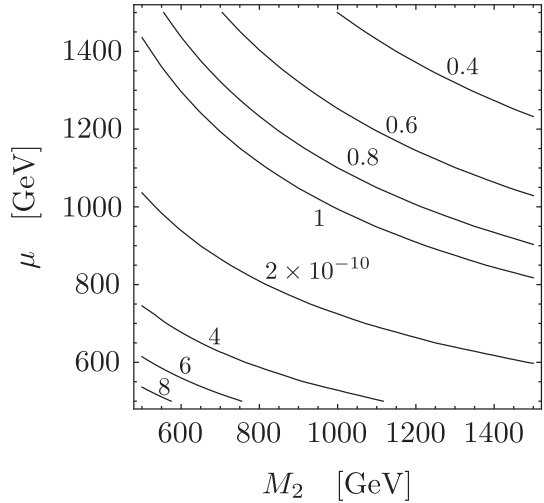}
  \caption{Contours of the difference between the full SUSY contribution
  to the muon $g-2$ and the ILC-reconstructed value, 
  $\delta a_\mu^{\rm (SUSY,th)} \equiv 
  a_\mu^{\rm (SUSY)}-a_\mu^{\rm (ILC)}$, on the
    Wino mass vs.~Higgsino mass plane.  }
  \label{fig:diff_gmin2}
\end{figure}

The reconstructed SUSY contribution $a_\mu^{\rm (ILC)}$
may not well
approximate the full contribution, $a_\mu^{\rm (SUSY)}$. 
The difference between $a_\mu^{\rm (SUSY)}$ and $a_\mu^{\rm (ILC)}$
comes from the unobserved neutralino and chargino contributions to 
the muon $g-2$, and should be understood as a theoretical error
in the reconstruction of $a_\mu^{\rm (SUSY)}$ in our procedure.
Let us define
\begin{align}
  \delta a_\mu^{\rm (SUSY,th)} \equiv
  a_\mu^{\rm (SUSY)} - a_\mu^{\rm (ILC)}.
\end{align}
This depends on $M_2$ and $\mu$ (as well as on the parameters listed
in Table \ref{table:accuracy}).  In Fig.~\ref{fig:diff_gmin2},
contours of constant $\delta a_\mu^{\rm (SUSY,th)}$ are shown for
$M_2>0$ with the underlying parameters in Table \ref{table:accuracy}.\footnote
{
We have checked that, when $M_2<0$, $|\delta
a_\mu^{\rm (SUSY,th)}|$ is smaller than that for $M_2>0$ with $|M_2|$
fixed.
}  
In particular, $\mu\tan\beta=6.1\times10^3\ {\rm GeV}$ is fixed.
Here, the uncertainties in the parameters listed in Table
\ref{table:accuracy} are omitted.  Obviously, $\delta a_\mu^{\rm
  (SUSY,th)}$ is suppressed as $M_2$ and $\mu$ become larger.  This is
because all the diagrams that contain Wino and Higgsino propagators
vanish in this limit, so that $a_\mu^{\rm (SUSY)}$ is well
approximated by the Bino--smuon diagram.  Thus, using lower bounds on
the Wino and Higgsino masses provided by collider experiments, a bound
on $\delta a_\mu^{\rm (SUSY,th)}$ can be obtained.  They will be
searched for effectively at LHC with $\sqrt{s}=13$ or $14\,{\rm
  TeV}$.\footnote{ Wino can be searched for by multi-lepton plus a
  large missing energy signature, while Higgsino can be by searches
  for multi-tau and/or standard model bosons together with a large
  missing energy.  } If the Wino and Higgsino masses are constrained
to be larger than $1\,{\rm TeV}$ ($1.5\,{\rm TeV}$) in future, $\delta
a_\mu^{\rm (SUSY,th)}$ is known to be smaller than $0.9 \times
10^{-10}$ ($0.3 \times 10^{-10}$) at our model point, which corresponds to $4\,\%$
($1\,\%$) of $a_\mu^{\rm (ILC)}$.  This is smaller than the dominant
error of the reconstruction of $a_\mu^{\rm (ILC)}$.

Finally we comment on higher order contributions to $a_\mu^{\rm(SUSY)}$.
Ref.~\cite{vonWeitershausen:2010zr} calculated
photonic SUSY two-loop corrections, which change the one-loop result 
by $\sim 10\,\%$.  They can be determined at ILC by the above procedure,
because all the parameters necessary for them are measured simultaneously. 
In this letter, they are neglected for simplicity, although it is
straightforward to include the contributions.  Also, corrections 
to the gaugino couplings and to the lepton Yukawa couplings in the 
left-right mixing parameters can be as large as $\sim 10\,\%$ \cite{Endo:2013lva,Fargnoli:2013zda,Marchetti:2008hw}.
Importantly, they are already taken into account in the reconstruction 
of $a_\mu^{\rm (ILC)}$. Most of the other two-loop contributions are 
considered to be suppressed in our sample point. However, electroweak 
and SUSY two-loop corrections to the SUSY one-loop diagrams, which have not been calculated, might be 
$\sim 10\,\%$ \cite{Stockinger:2006zn}.
Since they could be as large as the dominant error of the reconstruction,
it is important to calculate these two-loop contributions.

\section{Summary and Discussion}

In this letter, we have studied how and how accurately we can
reconstruct the SUSY contribution to the muon $g-2$ by using the
information available at ILC.  If $a_\mu^{\rm (SUSY)}$ is as large as
$2.6\times 10^{-9}$ to solve the muon $g-2$ anomaly, and also if all the
sleptons as well as the lightest neutralino are within the kinematical reach, 
ILC will be able to measure the MSSM parameters which are necessary to
estimate $a_\mu^{\rm (SUSY)}$.  We have discussed the procedures and
accuracies of their measurements.  It has been shown that, 
in the sample point we choose, the SUSY contribution to
the muon $g-2$ can be reconstructed with the uncertainty of 
$\sim 13\,\%$ at ILC with $\sqrt{s}=500\,{\rm GeV}$ and
an integrated luminosity ${\cal  L}\sim500\text{--}1000\,{\rm fb}^{-1}$.
This provides a very crucial test of the SUSY explanation to the  muon $g-2$ anomaly.

We should emphasize that the uncertainty depends on model points.
As we have shown, the dominant error in the reconstructed value of
$a_\mu^{\rm (ILC)}$ originates in the uncertainty of the left-right
mixing parameter $m_{\tilde\mu LR}^2$ in the sample point.  
For instance, if the heavier stau mass increases with the lighter one fixed, it is inferred from Eq.~\eqref{eq:sin2thl} that the reconstruction would be degraded.
On the contrary, if the charged sleptons in the second or third
generation are degenerate in masses, the determination of 
$\sin 2\theta_{\tilde{\mu}}$ could be improved considerably.
Unfortunately, slepton  productions have not been studied for ILC in such cases.

The present uncertainty of the experimental and SM values of 
the muon $g-2$ is about $30\,\%$ (see Eq.~\eqref{eq:deltaamu}).
Thus, the error in the reconstructed value of $a_\mu^{\rm (ILC)}$ 
is sub-dominant when we test the idea of solving the muon $g-2$ anomaly 
with the SUSY contribution.  However, the experimental measurement and 
theoretical calculation of the SM prediction will be improved 
in near future. The Fermilab experiment \cite{LeeRoberts:2011zz} 
and the J-PARC New $g-2$/EDM experiment \cite{Iinuma:2011zz} will reduce 
the experimental error at least by a factor $4\text{--}5$.
The uncertainty of the SM prediction is dominated by those in the hadronic 
contributions. They will be improved by experiments as well as lattice 
calculations. The uncertainty is expected to be reduced by a factor 
$2$ \cite{Hewett:2012ns}. As a result, if the experimental and SM central 
values would be unchanged, the error in $\Delta a_\mu$ could become 
as small as $\sim 10\,\%$, which is comparable to that in $a_\mu^{\rm (ILC)}$.
Then, a precise reconstruction of the SUSY contribution to the muon $g-2$
becomes crucial.

We made several assumptions to evaluate the uncertainty,
since we could not find enough information about the slepton production processes.
Precise studies of the slepton production process are strongly
recommended to deeply understand how useful ILC is to reconstruct
the SUSY contribution to the muon $g-2$.

\subsection*{Acknowledgements}
This work was supported by JSPS KAKENHI Grant Nos.~23740172 (M.E.),
22244021 (K.H. and T.M.), 25--1386 (S.I.), 22540263 (T.M.), and 23104001 (T.M.),
and also by the World Premier International Research Center
Initiative (WPI Initiative), MEXT, Japan.
The work of T.K. is partially supported by  Global COE Program
``the Physical Sciences Frontier", MEXT, Japan.

\providecommand{\href}[2]{#2}
\begingroup\raggedright

\endgroup

\begin{thebibliography}{99}

\bibitem{g-2_bnl2010}
{\bfseries Muon G-2} Collaboration,
 \href{http://dx.doi.org/10.1103/PhysRevD.73.072003}{Phys.\ Rev.\ {\bfseries D73} (2006) 072003}
{\ttfamily [\href{http://arxiv.org/abs/hep-ex/0602035}{hep-ex/0602035}]};
%
B.~L.~Roberts,
 \href{http://dx.doi.org/10.1088/1674-1137/34/6/021}{Chin.\ Phys.\ {\bfseries C34} (2010) 741--744}
{\ttfamily [\href{http://arxiv.org/abs/1001.2898}{arXiv:1001.2898}]}.

\bibitem{g-2_hagiwara2011}
K.~Hagiwara, A.~D.~Martin, D.~Nomura, and T.~Teubner,
 \href{http://dx.doi.org/10.1016/j.physletb.2007.04.012}{Phys.\ Lett.\ {\bfseries B649} (2007) 173--179}
{\ttfamily [\href{http://arxiv.org/abs/hep-ph/0611102}{hep-ph/0611102}]};
%
T.~Teubner, K.~Hagiwara, R.~Liao, A.~D.~Martin, and D.~Nomura,
 \href{http://dx.doi.org/10.1088/1674-1137/34/6/019}{Chin.\ Phys.\ {\bfseries C34} (2010) 728--734}
{\ttfamily [\href{http://arxiv.org/abs/1001.5401}{arXiv:1001.5401}]};
%
K.~Hagiwara, R.~Liao, A.~D.~Martin, D.~Nomura, and T.~Teubner,
 \href{http://dx.doi.org/10.1088/0954-3899/38/8/085003}{J.\ Phys.\ G {\bfseries G38} (2011) 085003}
{\ttfamily [\href{http://arxiv.org/abs/1105.3149}{arXiv:1105.3149}]}.

\bibitem{g-2_davier2010}
M.~Davier, {\em et al.},
 \href{http://dx.doi.org/10.1140/epjc/s10052-009-1219-4}{Eur.\ Phys.\ J.\ {\bfseries C66} (2010) 127--136}
{\ttfamily [\href{http://arxiv.org/abs/0906.5443}{arXiv:0906.5443}]};
%
M.~Davier, A.~Hoecker, B.~Malaescu, C.~Z.~Yuan, and Z.~Zhang,
 \href{http://dx.doi.org/10.1140/epjc/s10052-010-1246-1}{Eur.\ Phys.\ J.\ {\bfseries C66} (2010) 1--9}
{\ttfamily [\href{http://arxiv.org/abs/0908.4300}{arXiv:0908.4300}]};
%
M.~Davier, A.~Hoecker, B.~Malaescu, and Z.~Zhang,
 \href{http://dx.doi.org/10.1140/epjc/s10052-010-1515-z}{Eur.\ Phys.\ J.\ {\bfseries C71} (2011) 1515} {\ttfamily
  [\href{http://arxiv.org/abs/1010.4180}{arXiv:1010.4180}]}.
Erratum \href{http://dx.doi.org/10.1140/epjc/s10052-012-1874-8}{Ibid.
  {\bfseries C72} (2012) 1874}.

\bibitem{Aoyama:2012wk} 
  T.~Aoyama, M.~Hayakawa, T.~Kinoshita and M.~Nio,
  Phys.\ Rev.\ Lett.\  {\bf 109}, 111808 (2012)
  [arXiv:1205.5370 [hep-ph]].

\bibitem{Gnendiger:2013pva} 
  C.~Gnendiger, D.~St\"ockinger and H.~St\"ockinger-Kim,
  Phys.\ Rev.\ D {\bf 88}, 053005 (2013)
  [arXiv:1306.5546 [hep-ph]].


\bibitem{Lopez:1993vi} 
  J.~L.~Lopez, D.~V.~Nanopoulos and X.~Wang,
  Phys.\ Rev.\ D {\bf 49}, 366 (1994)
   {\ttfamily [\href{http://arxiv.org/abs/hep-ph/9308336}{hep-ph/9308336}]}.

\bibitem{Chattopadhyay:1995ae} 
  U.~Chattopadhyay and P.~Nath,
  Phys.\ Rev.\ D {\bf 53}, 1648 (1996)
   {\ttfamily [\href{http://arxiv.org/abs/hep-ph/9507386}{hep-ph/9507386}]}.

\bibitem{Moroi:1995yh} 
  T.~Moroi,
  Phys.\ Rev.\ D {\bf 53}  (1996) 6565
  [Erratum-ibid.\ D {\bf 56}  (1997) 4424]
  {\ttfamily [\href{http://arxiv.org/abs/hep-ph/9512396}{hep-ph/9512396}]}.

\bibitem{Behnke:2013xla} 
  T.~Behnke, J.~E.~Brau, B.~Foster, J.~Fuster, M.~Harrison, J.~M.~Paterson, M.~Peskin and M.~Stanitzki {\it et al.},
  {\ttfamily [\href{http://arxiv.org/abs/1306.6327}{arXiv:1306.6327}]}.

\bibitem{Endo:2013lva} 
  M.~Endo, K.~Hamaguchi, T.~Kitahara and T.~Yoshinaga,
  JHEP  (2013) in press.
   {\ttfamily [\href{http://arxiv.org/abs/1309.3065}{arXiv:1309.3065}]}.

\bibitem{Hikasa:1995bw} 
  K.~-i.~Hikasa and Y.~Nakamura,
  Z.\ Phys.\ C {\bf 70}, 139 (1996)
  [Erratum-ibid.\ C {\bf 71}, 356 (1996)]
  {\ttfamily [\href{http://arxiv.org/abs/hep-ph/9501382}{hep-ph/9501382}]}.

\bibitem{Nojiri:1996fp} 
  M.~M.~Nojiri, K.~Fujii and T.~Tsukamoto,
  Phys.\ Rev.\ D {\bf 54}, 6756 (1996)
   {\ttfamily [\href{http://arxiv.org/abs/hep-ph/9606370}{hep-ph/9606370}]}.

\bibitem{Cheng:1997sq} 
  H.~-C.~Cheng, J.~L.~Feng and N.~Polonsky,
  Phys.\ Rev.\ D {\bf 56}, 6875 (1997)
  {\ttfamily [\href{http://arxiv.org/abs/hep-ph/9706438}{hep-ph/9706438}]}; 
  Phys.\ Rev.\ D {\bf 57}, 152 (1998)
  {\ttfamily [\href{http://arxiv.org/abs/hep-ph/9706476}{hep-ph/9706476}]}; 
  E.~Katz, L.~Randall and S.~-f.~Su,
  Nucl.\ Phys.\ B {\bf 536}, 3 (1998)
   {\ttfamily [\href{http://arxiv.org/abs/hep-ph/9801416}{hep-ph/9801416}]}.

\bibitem{Nojiri:1997ma} 
  M.~M.~Nojiri, D.~M.~Pierce and Y.~Yamada,
  Phys.\ Rev.\ D {\bf 57}, 1539 (1998)
  {\ttfamily [\href{http://arxiv.org/abs/hep-ph/9707244}{hep-ph/9707244}]}.

\bibitem{Fargnoli:2013zda} 
  H.~Fargnoli, C.~Gnendiger, S.~Passehr, D.~St\"ockinger and H.~St\"ockinger-Kim,
  arXiv:1309.0980 [hep-ph].

\bibitem{AguilarSaavedra:2005pw} 
  J.~A.~Aguilar-Saavedra, A.~Ali, B.~C.~Allanach, R.~L.~Arnowitt, H.~A.~Baer, J.~A.~Bagger, C.~Balazs and V.~D.~Barger {\it et al.},
  Eur.\ Phys.\ J.\ C {\bf 46}, 43 (2006)
  {\ttfamily [\href{http://arxiv.org/abs/hep-ph/0511344}{hep-ph/0511344}]}.

\bibitem{ATLAS2013049}
{\bfseries ATLAS} Collaboration,
 \href{http://cdsweb.cern.ch/record/1547565}{ATLAS--CONF--2013--049} (2013).

\bibitem{CMSPASSUS13006}
{\bfseries CMS} Collaboration,
 \href{http://cds.cern.ch/record/1563142}{CMS PAS SUS--13--006} (2013).

\bibitem{PDG}
J. Beringer et al. (Particle Data Group), Phys. Rev. D86, 010001 (2012) and 2013 partial update for the 2014 edition. 

\bibitem{ATLAS2013035}
{\bfseries ATLAS} Collaboration,
 \href{http://cdsweb.cern.ch/record/1532426}{ATLAS--CONF--2013--035} (2013).

\bibitem{Bechtle:2009em} 
  P.~Bechtle, M.~Berggren, J.~List, P.~Schade and O.~Stempel,
  Phys.\ Rev.\ D {\bf 82}, 055016 (2010)
  {\ttfamily [\href{http://arxiv.org/abs/0908.0876}{arXiv:0908.0876}]}.

\bibitem{Tsai:1971vv} 
  Y.~-S.~Tsai,
  Phys.\ Rev.\ D {\bf 4}, 2821 (1971)
  [Erratum-ibid.\ D {\bf 13}, 771 (1976)];
  T.~Hagiwara, S.~-Y.~Pi and A.~I.~Sanda,
  Annals Phys.\  {\bf 106}, 134 (1977);
  H.~-K.~Kuhn and F.~Wagner,
  Nucl.\ Phys.\ B {\bf 236}, 16 (1984).

\bibitem{Nojiri:1994it} 
  M.~M.~Nojiri,
  Phys.\ Rev.\ D {\bf 51}, 6281 (1995)
   {\ttfamily [\href{http://arxiv.org/abs/hep-ph/9412374}{hep-ph/9412374}]}.

\bibitem{Grannis:2002xd} 
  P.~D.~Grannis,
   {\ttfamily [\href{http://arxiv.org/abs/hep-ex/0211002}{hep-ex/0211002}]}.

\bibitem{Baer:2013cma} 
  H.~Baer, T.~Barklow, K.~Fujii, Y.~Gao, A.~Hoang, S.~Kanemura, J.~List and H.~E.~Logan {\it et al.},
  {\ttfamily [\href{http://arxiv.org/abs/1306.6352}{arXiv:1306.6352}]}.

\bibitem{Boos:2003vf} 
  E.~Boos, H.~U.~Martyn, G.~A.~Moortgat-Pick, M.~Sachwitz, A.~Sherstnev and P.~M.~Zerwas,
  Eur.\ Phys.\ J.\ C {\bf 30}, 395 (2003)
  {\ttfamily [\href{http://arxiv.org/abs/hep-ph/0303110}{hep-ph/0303110}]}.

\bibitem{AguilarSaavedra:2001rg} 
  J.~A.~Aguilar-Saavedra {\it et al.}  [ECFA/DESY LC Physics Working Group Collaboration],
  {\ttfamily [\href{http://arxiv.org/abs/hep-ph/0106315}{hep-ph/0106315}]}.

\bibitem{Tsukamoto:1993gt} 
  T.~Tsukamoto, K.~Fujii, H.~Murayama, M.~Yamaguchi and Y.~Okada,
  Phys.\ Rev.\ D {\bf 51}, 3153 (1995).

\bibitem{Martyn:2004ew} 
  H.~-U.~Martyn,
    {\ttfamily [\href{http://arxiv.org/abs/hep-ph/0406123}{hep-ph/0406123}]}.

\bibitem{Freitas:2004re} 
  A.~Freitas, H.~-U.~Martyn, U.~Nauenberg and P.~M.~Zerwas,
  {\ttfamily [\href{http://arxiv.org/abs/hep-ph/0409129}{hep-ph/0409129}]}.

\bibitem{Allanach:2002nj} 
  B.~C.~Allanach, M.~Battaglia, G.~A.~Blair, M.~S.~Carena, A.~De Roeck, A.~Dedes, A.~Djouadi and D.~Gerdes {\it et al.},
  Eur.\ Phys.\ J.\ C {\bf 25}, 113 (2002)
  {\ttfamily [\href{http://arxiv.org/abs/hep-ph/0202233}{hep-ph/0202233}]}.

\bibitem{Brau:2012hv} 
  J.~E.~Brau, R.~M.~Godbole, F.~R.~L.~Diberder, M.~A.~Thomson, H.~Weerts, G.~Weiglein, J.~D.~Wells and H.~Yamamoto,
   {\ttfamily [\href{http://arxiv.org/abs/1210.0202}{arXiv:1210.0202}]}.

\bibitem{Martyn:2004jc} 
  H.~-U.~Martyn,
  {\ttfamily [\href{http://arxiv.org/abs/hep-ph/0408226}{hep-ph/0408226}]}.

\bibitem{Berggren:2009uj} 
  M.~Berggren, N.~d'Ascenzo, P.~Schade and O.~Stempel,
    {\ttfamily [\href{http://arxiv.org/abs/0902.2434}{arXiv:0902.2434}]}.

\bibitem{Freitas:2002gh} 
  A.~Freitas, J.~Kalinowski, B.~Ananthanarayan, A.~Bartl, G.~A.~Blair, C.~Blochinger, E.~Boos and A.~Brandenburg {\it et al.},
  {\ttfamily [\href{http://arxiv.org/abs/hep-ph/0211108}{hep-ph/0211108}]}.

\bibitem{Freitas:2003yp} 
  A.~Freitas, A.~von Manteuffel and P.~M.~Zerwas,
  Eur.\ Phys.\ J.\ C {\bf 34}, 487 (2004)
   {\ttfamily [\href{http://arxiv.org/abs/hep-ph/0310182}{hep-ph/0310182}]}.

\bibitem{Kilian:2006he} 
  W.~Kilian and P.~M.~Zerwas,
   {\ttfamily [\href{http://arxiv.org/abs/hep-ph/0601217}{hep-ph/0601217}]}.

\bibitem{Weiglein:2004hn} 
  G.~Weiglein {\it et al.}  [LHC/LC Study Group Collaboration],
  Phys.\ Rept.\  {\bf 426}, 47 (2006)
  {\ttfamily [\href{http://arxiv.org/abs/hep-ph/0410364}{hep-ph/0410364}]}.

\bibitem{MoortgatPick:2005cw} 
  G.~Moortgat-Pick {\it et al.},
  Phys.\ Rept.\  {\bf 460}, 131 (2008)
  {\ttfamily [\href{http://arxiv.org/abs/hep-ph/0507011}{hep-ph/0507011}]}.

\bibitem{vonWeitershausen:2010zr}
  P.~von Weitershausen, M.~Schafer, H.~St\"ockinger-Kim and D.~St\"ockinger,
  Phys.\ Rev.\ D {\bf 81} (2010) 093004
   {\ttfamily [\href{http://arxiv.org/abs/1003.5820 }{arXiv:1003.5820}]}.

\bibitem{Marchetti:2008hw} 
  S.~Marchetti, S.~Mertens, U.~Nierste and D.~St\"ockinger,
  Phys.\ Rev.\ D {\bf 79}, 013010 (2009)
  [arXiv:0808.1530 [hep-ph]].

\bibitem{Stockinger:2006zn} 
  D.~St\"ockinger,
  J.\ Phys.\ G {\bf 34}, R45 (2007)
    {\ttfamily [\href{http://arxiv.org/abs/hep-ph/0609168}{hep-ph/0609168}]}.

\bibitem{LeeRoberts:2011zz}
  B.~Lee Roberts [Fermilab P989 Collaboration],
  Nucl.\ Phys.\ Proc.\ Suppl.\  {\bf 218} (2011) 237.

\bibitem{Iinuma:2011zz}
  H.~Iinuma [J-PARC New g-2/EDM experiment Collaboration],
  J.\ Phys.\ Conf.\ Ser.\  {\bf 295} (2011) 012032.

\bibitem{Hewett:2012ns}
  See, for example, J.~L.~Hewett {\it et al.},
  {\ttfamily [\href{http://arxiv.org/abs/1205.2671}{arXiv:1205.2671}]}.

\end{thebibliography}
\end{document}